\documentclass[journal]{IEEEtran}
\usepackage{xcolor,soul,framed} 
\usepackage{cite}
\usepackage{multirow}
\usepackage{fontawesome}
\colorlet{shadecolor}{yellow}
\usepackage[pdftex]{graphicx}
\graphicspath{{../pdf/}{../jpeg/}}

\usepackage{amsmath}
\usepackage{amsthm}
\DeclareGraphicsExtensions{.pdf,.jpeg,.png}
\usepackage{caption}
\usepackage{subcaption}
\usepackage[colorinlistoftodos]{todonotes}
\usepackage{comment}
\usepackage{bm}
\usepackage{enumerate}

 \newcommand\quickthings[1]{}
\newcommand\maybelater[1]{}
\newcommand\peter[1]{}
\usepackage[linesnumbered,ruled,vlined]{algorithm2e}
\begin{document}
\title{Exploring adversarial attacks in federated learning for medical imaging}

\author{
    Erfan~Darzi$^{1,3*}$, 
    Florian Dubost$^{2}$, 
    N.M. Sijtsema$^{3}$, 
    P.M.A van Ooijen$^{3}$, 
    \\
    \textit{$^{1}$Harvard University, Boston, MA, USA} \\
    \textit{$^{2}$Google, Mountain View,  CA, USA}\\
    \textit{$^{3}$Department of Radiotherapy, University Medical Center Groningen, \\University of Groningen, Groningen, The Netherlands}\\
}

\bstctlcite{IEEEexample:BSTcontrol} 

\maketitle
\maketitle

\begin{abstract}


Federated learning offers a privacy-preserving framework for medical image analysis but exposes the system to adversarial attacks. This paper aims to evaluate the vulnerabilities of federated learning networks in medical image analysis against such attacks. Employing domain-specific MRI tumor and pathology imaging datasets, we assess the effectiveness of known threat scenarios in a federated learning environment. Our tests reveal that domain-specific configurations can increase the attacker's success rate significantly. The findings emphasize the urgent need for effective defense mechanisms and suggest a critical re-evaluation of current security protocols in federated medical image analysis systems.

\end{abstract}


%
\IEEEpeerreviewmaketitle



\section{Introduction}

\IEEEPARstart{F}ederated learning is a machine learning paradigm that enables hospitals and other healthcare providers to train machine learning models jointly without sharing sensitive data. This solves data governance concerns and is applied to a variety of medical applications, including brain tumor classification and segmentation, breast density classification, and covid-19 detection\cite{darzidehkalani2023comparative}. Although federated learning has mitigated risks and concerns about multi-institutional collaborations in healthcare, it also exposes hospitals to more security and privacy threats originating from outside the secured hospital domain\cite{darzidehkalani2022federated1,liu2022threats}.

\begin{figure}[t!]
\centering
\includegraphics[width=0.49\textwidth]{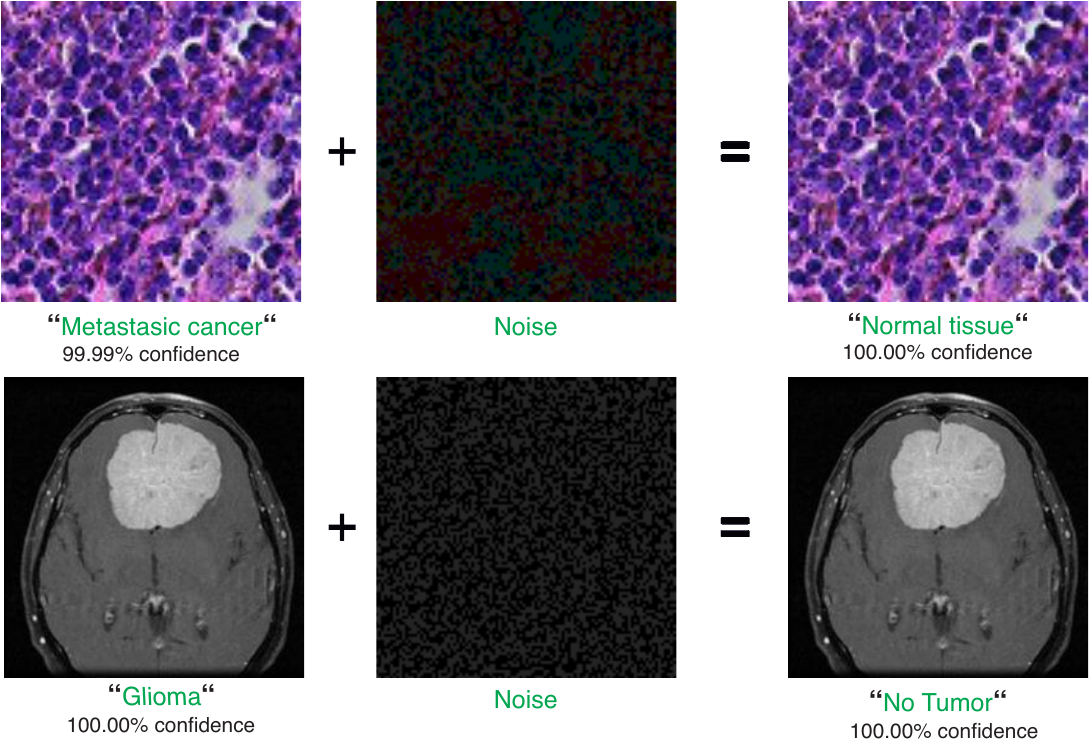} 
\caption{A schema of adversarial attacks to cancer detection systems}
\label{figure:schema}
\end{figure}

 For instance, a client may attempt to deceive the global model and other clients into ignoring or misclassifying a specific class of tumors by deliberately adding fake samples or altering labels of their training data, as seen in backdoor attacks. Sensitive information from other hospitals could be obtained by decoding the models and inverting the received gradients. Figure \ref{figure:schema} shows a schema of adversarial attacks on cancer detection systems. An attacker could potentially reconstruct sensitive patient images from the gradients, thereby compromising the privacy of the patients involved. 

 Such issues exist in healthcare, and there are specific incentives for such actions. The large healthcare economy causes more benefit for malicious behavior, as there are existing reports of pervasive fraud in healthcare. Decoding the model and stealing sensitive information can result in financial gains for the attackers; they can sell the private information. Being able to control and change the model's prediction on specific cases allows a malicious client to fool any hospital or insurance company that uses that model. 

 Consider an attacker who has access to and can modify a tumor detection model's prediction. This may be used to approve a tumor case from an insurance company that uses AI-based algorithms to approve a disease and receive a refund. Medical record manipulation for financial purposes \cite{ma2021understanding} fake trial reports\cite{george2015data} in radiology \cite{chowdhry2014image} and pathology\cite{suvarna2001histopathology} are existing practices, and already made billions of dollars profit \cite{graese2016assessing} for fraudsters.  

Some methods have been developed to protect federated learning networks from manipulations. One example is differential privacy, where all model parameters are truncated and Gaussian noise is added in each round of communication and is shown to be effective against privacy attacks. Another example is secure aggregation, where communicated information is encrypted using a cryptographic protocol. For backdoor and poisoning attacks, where a client aims to change a global model prediction for certain classes by using fake training data or by corrupting the local updates, there are some statistical tests that can spot unusual client updates. More sophisticated optimization-based methods are also devised to protect the models from such attacks \cite{xiao2022sca}.

 In this paper, we focus on the challenges and threats faced by federated learning in medical imaging domain, discussing the security and privacy risks related to malicious clients, data manipulation, and potential financial incentives for attackers. 

\subsection{Adversarial attacks}

 Adversarial attacks are caused by the linear characteristics of high-dimensional data \cite{goodfellow2014explaining}. In federated settings, these attacks are also known as evasion attacks\cite{biggio2013evasion} and are defined as if a malicious client tries to manipulate other clients with fake data. Attack transferability is the ability of an adversarial attack to be successful against unseen models. 

We investigate adversarial attacks in the scope of federated learning and medical image analysis systems. Federated learning vulnerabilities are known to occur on top of the vulnerabilities of a centralized setting like the traditional ML vulnerabilities. The unique vulnerabilities of federated learning are: \textit{More data}: Each model received in every round has updated information about other clients. Adversaries can exploit this information to prepare stronger attacks\cite{sun2019can}. \textit{Untrusted environments}: In federated settings, the training process involves a lot of participants. Among them, one adversary could act maliciously. \textit{Standardized data pipelines}: Data standardization steps are inherent in many federated learning networks in hospitals. Standardization process will cause clients to have more similar local data distributions. Similarity in data leads to higher transferability between models \cite{van2022ai}.





Medical images are more vulnerable to adversarial attacks than natural images for two major reasons: \textit{Feature representation}: Medical images have a narrower high-dimensional feature representation of images than natural images, causing trained networks to be over-parameterized \cite{ma2021understanding}. Over-parameterized networks are inherently easier to fool\cite{Ye_2019_ICCV}. \textit{Unique texture}. In addition, medical images have limited texture diversity, and small texture perturbation in medical images can confuse classifiers to a high degree. These features can be of advantage to the adversary’s attacks. For example, an adversary can perturb the texture in irrelevant areas and fool the classifier without manipulating the more important parts, such as the tumor area\cite{ma2021understanding,finlayson2019adversarial,gupta2022vulnerability}.

\subsection{Transferability factors}

 The research is ongoing on attack transferability\cite{gao2022boosting}. In the medical image analysis domain, transferability analysis has been done in a centralized ML setting, and the results are found to be domain specific. Factors such as data disparity, perturbation degree, and pre-training are shown to be crucial in attack transferability.
However, the extent and optimal values of these factors may vary significantly. This variation can be attributed to the target imaging domain's characteristics, image texture, and the adversary's knowledge. In a federated setting characterized by a higher level of complexity, findings from centralized settings may have limited relevance. In this study, our primary focus is on evaluating the potential factors that might influence attack transferability in a federated learning environment. We investigate three factors: perturbation degree, attack step parameter, and the domain of attack, and discuss the impact degree of each factor on attack transferability.
 
Unlike research focused on natural images, we find that the type of medical imaging modality, such as MRI or pathology, significantly impacts the transferability of these attacks. This is due to the unique texture and resolution of medical images, adding a layer of complexity that does not exist in studies involving natural images. In addition, our paper is one of the first to evaluate how various factors like perturbation degree and attack step parameters influence the transferability of attacks in a federated learning healthcare context.  
\section{Preliminaries and related works}
\label{sec:prelimianries}
\subsection{Federated Learning}

Federated learning enables multiple data owners with private datasets to jointly train a global model based on local models. The  optimization problem could be formulated as
${w} = \sum\limits_{i=1}^{N}{p_{i}{w}_{i}}$ , ${w}_{i}=\arg\min\limits_{{w}_{i}}{\left(\mathcal{L}(\mathcal{D}_{i};{w})\right)}$ where $N$ is the number of data owners, $\mathcal{L}(\mathcal{D}_{i};{w})$ is a loss function indicating global model parameters  ${w}$ of local datasets.  
 The learning procedure is an iterative process containing local and global steps. Each data owner trains a model received from a global server on its local dataset in local iterations. The global server updates the global model by aggregating the updated local models. Then it sends it back to clients for the next round. 
 
 The global server selects a subset of clients at each global round and sends the most recent global model to them. Then each client performs local training over its dataset for a selected number of epochs. The updated local models are calculated on selected batches. Local optimization can be formulated as ${w}_i \leftarrow {w}-\eta\cdot \nabla \mathcal{L}({w};\mathcal{D}_{i})$, where 
 $\eta$ is the learning rate. Several local iterations might be required to go over all the local data. Local training procedures can be done for several local epochs.
 The global model can be updated based on the local models ${w}_i$ and is shared for aggregation: 
${w} = \sum\limits_{i=1}^{N}{p_{i}{w}_{i}}$, to update the global model for the next federated round.

Differential privacy (DP) requires federated learning parties to ensure an attacker can not distinguish data records. For multi-party systems $\mathcal M: \mathcal{X}\rightarrow \mathcal{R}$ mapping from domain $\mathcal{X}$ to target domain $\mathcal{R}$, differential privacy  $(\epsilon, \delta)$-DP defines a measure to evaluate  performance of privacy preserving mechanisms.
For two adjacent datasets $\mathcal D_i, \mathcal D_i'$. DP introduces a bounding parameter $\epsilon > 0$ , representing the ratio of probabilities of two datasets bounded by performing the privacy preserving mechanism  $\delta$. It can be summarized by the following definition \cite{dwork2014algorithmic}:
 
Mapping $\mathcal M: \mathcal{X}\rightarrow \mathcal{R}$ is $(\epsilon, \delta)$-DP,
if for all measurable subsets of target domain $\mathcal S\subseteq \mathcal{R}$ and for any two adjacent datasets $\mathcal D_i, \mathcal D_i'\in \mathcal{X}$, 
\begin{equation}\label{equ:Differential privacy}
\emph{Pr}[\mathcal M(\mathcal D_i)\in \mathcal S]\leq e^{\epsilon}\emph{Pr}[\mathcal M(\mathcal D_i')\in \mathcal S]+\delta.
\end{equation}
In FL setting,  $(\epsilon, \delta)$-DP can be achieved by adding noise to the updated models.
Global differential privacy is a privacy mechanism that imposes a double-sided $(\epsilon, \delta)$-DP requirement for both uplink and downlink channels\cite{wei2020federated}.
From the uplink perspective, all clients $1\leq i\leq N$, clip their updates  $\Vert{w}_{i}\Vert \leq C$, where ${w}_{i}$ denotes the updated weights from the $i$-th client before perturbation and $C$ is the clipping threshold. To satisfy a $(\epsilon, \delta)$-DP requirement for the downlink channels, additional noise ${n}_{\text i}$ is   added by the server, so each client $i$ receives a $\tilde{w_i}$ perturbed model .




\subsection{Adversarial attacks }

Introduced by \cite{szegedy2013intriguing}, 
adversarial examples are feature space manipulations with respect to linear decision boundaries,
\cite{goodfellow2014explaining} and are categorized based on the adversary's goal and knowledge.
\\\textbf{Adversary's goal} Adversarial attacks can be categorized based on the adversary's goal. \textit{Untargetted attacks} aim to reduce the model performance,  regardless of the class to which a test sample belongs. \textit{Targeted attacks} force the model to output certain labels.
\\\textbf{Adversary's knowledge} Based on the adversary's knowledge, the attack can be  \textit{white-box},
meaning that the adversary has complete knowledge about other clients' network architecture, gradients, and parameters. In such a setting, it can easily manipulate the model. White-box attacks have been extensively investigated in the literature\cite{xu2020adversarial}. In \textit{black-box} heuristics, the adversary does not have access to the target model. In the black-box setting, however, it can interact with the model. Adversary can feed inputs and receive outputs of the target model and improve the attack by observing the model outputs. Some black box might have limited knowledge about design of the target model.
\\Popular attack methods are Projected Gradient Descent (PGD), Fast Gradient Sign Method (FGSM), and Basic Iterative Method (BIM). Each will be discussed in this section.

\textbf{FGSM:} Fast Gradient Sign Method (FGSM)   is a fast yet effective method which produces adversary images with one step of calculation. Assuming input $x$ and its corresponding target $t$, FGSM calculates gradient of $x$ with respect to the loss function ${\partial \mathcal{L}}/ \partial {x}$.

\begin{equation}
\label{eqt:fgsm}
    \bm{\hat{x}} = \bm{x} + \epsilon \cdot sgn\big(\nabla_{\bm{x}}{\mathcal{L}}(g(\bm{x};w)\big)
\end{equation}

where epsilon $\epsilon$ is a hyper-parameter which determines adversarial noise level. $g(\bm{x};\bm{w})$ is the output of neural network with respect to the input $x$ and parameter set $w$. $sgn(.)$ is the sign function. The result of sign function goes through a clipping function to impose a maximum bound in change to the perturbation $\epsilon \cdot sgn\big(\nabla_{\bm{x}}{\mathcal{L}}(g(\bm{x};\bm{w})\big)\in [-1,1]$.

\textbf{BIM:}  Basic Iterative Method (BIM) is an extension of the FGSM method, proposed by Kurakin et al.\cite{kurakin2018adversarial}. BIM repeatedly performs the FGSM process, using a small step-size and $\bm{\hat{x}^{1}} = \bm{x}$. BIM is stronger than FGSM and requires smaller perturbations.

\textbf{PGD:} Madry et al.  proposed their own version of the BIM attack. In Projected Gradient Descent (PGD)
the attack starts with a uniform random initialization. The update formula for PGD attack can be written as: 
\begin{equation}
\label{eqt:pgd}
    \bm{\hat{x}}^{t+1}=\Pi_{P_\epsilon(\bm{x})} \Big( \bm{\hat{x}}^t + \alpha \cdot sgn\big(\nabla_{\bm{x}}{\mathcal{L}}(g(\bm{\hat{x}}^t;\bm{w}),t)\big)\Big)
\end{equation}
Where $\bm{\hat{x}}^{k}$ is the perturbed data in $k$-th iteration, and $P_\epsilon(\bm{x})$ is the projected gradient descent function, which is done by first finding sign values and then projecting the result to a small neighborhood of the input ${x}$. These possible parameter spaces determine the set of PGD attack samples that an adversary can use. PGD is one of the strongest attacks and is a universal first-order adversarial method. Technically, PGD is similar to BIM but with a small random initialization, aside from formulating the problem as a projected gradient. We report results with PGD. With equal iterations, BIM had the same results, or the difference was negligible.
\subsection{Medical imaging domain}
\label{sec:related}

The effect of adversarial attacks on medical image analysis has been studied in several works. Tasks such as chest X-ray, MRI \cite{ma2021understanding} and CT scan segmentation and classification are vulnerable to adversarial attacks. Several studies were performed to discover important parameters related to attack transferability. Gradients of different samples in one batch, extent of data augmentation , variation in input gradients have been shown to be important in transferability. Zhang et al. analyzed the impact of transfer learning methods on black-box attack\cite{zhang2019theoretically} .
 Another work utilized the transferability of adversarial examples to enhance  robustness, although their method was proposed for a white-box setting \cite{zhou2018transferable}. Model aggregation mechanism has also been studied as a point of attack. For instance, the work by Xiao et al. \cite{xiao2022sca} delves into vulnerabilities present during model aggregation in federated learning, providing a framework to understand sybil-based attacks. This research exemplifies the types of security vulnerabilities that can be applied to our focus on medical imaging.

In medical image analysis,  perturbation degree
has been shown previously as a less explored but highly deterministic parameter in attack setting, and might need visual tuning . Optimal values of standard black-box, and white-box are domain-specific. Also iteration steps $\alpha$, can be highly deterministic in centralized setting . 

Methods to detect the attack or defend the models from adversaries were also discussed in the ML domain. Adding noise to the exchanged model with clipping model updates is effective in several forms of adversarial attacks . In norm-bound defense, the server enforces an upper-limit norm-bound. Several studies have investigated black-box PGD attacks in a federated environment with norm-bound situation \cite{sun2019can }. 

In the clinical setting, \cite{shao2019stochastic} DP models are used in clinical EHR data and neuroimaging data  in multi-site setting.
In synthesizing the aforementioned literature, several important observations can be made. Firstly, the focus on medical imaging tasks such as MRI, CT scans, and chest X-rays varies in its approach to adversarial attacks. For instance, the work by Ma et al. \cite{ma2021understanding} largely dwells on the vulnerabilities in medical imaging but doesn't explore the issue of black-box transferability, which Zhang et al. \cite{zhang2019theoretically} delves into but in a more generic context. While Zhang's work underscores the role of transfer learning in black-box attacks, it may lack the specialized focus that a medical imaging setting requires.

Secondly, when it comes to defense mechanisms, the literature seems to be fragmented. Some studies focus on norm-bound defenses or adding noise to counter adversarial attacks, but these approaches may not be universally effective across various medical imaging domains. For instance, the work of Sun et al. \cite{sun2019can} investigates PGD attacks within a federated environment but leaves the applicability in medical imaging largely unexplored. Another noteworthy approach in the medical imaging domain is watermarking. As proposed by Hosny et al., watermarking methods offer a high level of visual imperceptibility and robustness against various kinds of attacks including geometric distortions and signal processing attacks \cite{hosny2018parallel} . However, these watermarking methods are yet to be validated against the range of adversarial attacks that could occur in a federated environment, leaving an open question that our paper seeks to address.

Lastly, the majority of these studies consider either a federated or a centralized setup, but a comparative analysis between these configurations in the context of medical imaging is missing. Studies like those by Zhou et al. \cite{zhou2018transferable} that discuss the transferability of adversarial examples for robustness are generally confined to white-box settings and don't extend their findings to a federated scenario.

\section{Threat model}
\label{sec:ourattack}
In our federated learning environment, we propose an attack scenario where an individual client becomes malicious or is compromised by an external adversary. Contrary to the general assumption that federated learning operates in an honest-but-curious setting, we focus on the case where a malicious client aims to intentionally deceive the global model by  generating adversarial samples similar to its local real data. 
\\\textbf{Goal of adversary:}
The adversary's goal is to manipulate the fully trained global model so that the global model has a higher classification error.  \\
\textbf{Knowledge of adversary:}
In a realistic scenario, the malicious party only can see its own data $D_{i}$, and knows the neural network's architecture and the global model weights that it receives each round. This grants the adversary partial knowledge, enough to execute internal adversarial attacks.
\begin{algorithm}[t]
\SetAlgoLined
\SetKwInOut{Input}{Input}
\SetKwInOut{Output}{Output}

\Input{$N$, $K$, $T$, $C$, $\eta$, $\epsilon$, $\delta$, $attack$}
\Output{Trained global model $w^T$}

\BlankLine

Initialize the global model $w^0$\;

\For{$t=1,\dots,T$}{
  Sample $m$ clients from $N$ clients without replacement\;
  
  \For{each selected client $i$}{
    Client $i$ receives the global model $w^{t-1}$\;
    Clip client $i$'s update: $\Vert{w}_{i}\Vert \leq C$\;
    Perform local training with DP: $w_i^t \leftarrow \text{LocalUpdate}(w^{t-1}, \mathcal{D}_i, C, \eta, \epsilon, \delta)$\;
  }
  
  Aggregate local models: $w^t = \sum\limits_{i=1}^{m}\frac{|\mathcal{D}_i|}{\sum_{i=1}^m |\mathcal{D}_i|} w_i^t$\;
}

\uIf{$attack = \text{PGD}$}{
  Evaluate the global model on the test dataset using PGD attack\;
}
\Else{
  Evaluate the global model on the test dataset using FGSM attack\;
}

\Return $w^T$\;

\caption{Federated Learning with DP and Test-time PGD/FGSM Attacks}
\label{alg:FLDPAttack}
\end{algorithm}

\textbf{Capability of adversary:} 
No special privilege or capability is assumed for the adversary. Similar to other clients, the adversary can not manipulate other clients' data or the general learning process. (e.g., local computations, communication with central server and aggregation process, DNN architecture, and optimization functions). It can interact with the model, query the outputs, and calculate the gradients.
The adversary uses its training data for federated training, and test data to perform adversarial attacks.  Using this capability, the adversary employs its locally-generated test data to perform calculated adversarial attacks during the test or inference phase.


\section{Experimental setup}
\begin{figure*}[t!]
 \centering
 \includegraphics[width=0.8\textwidth]{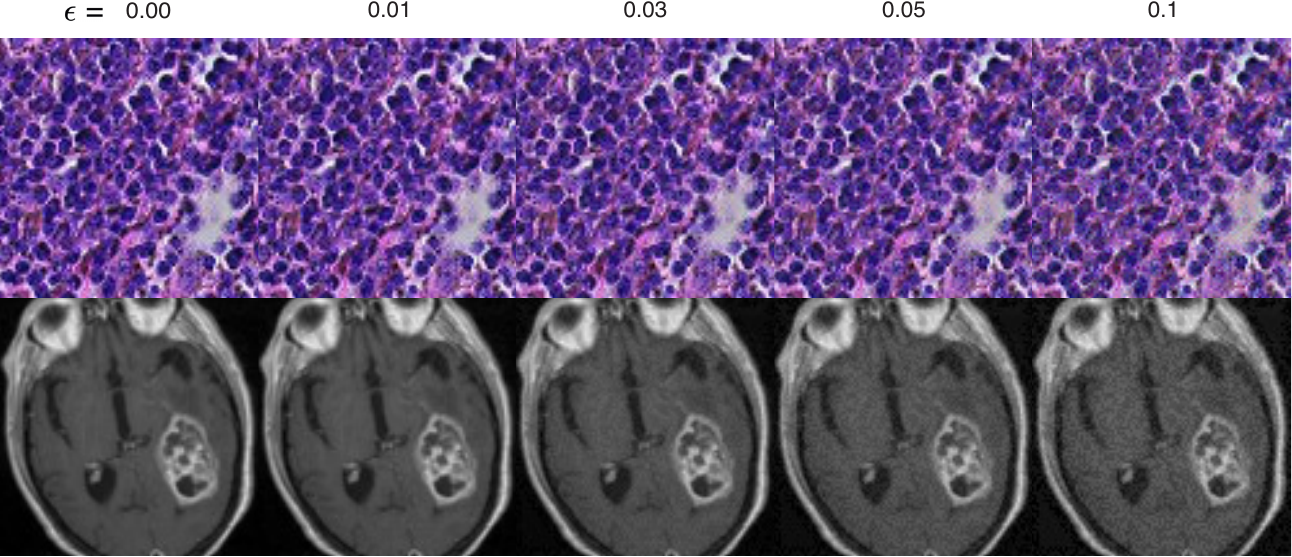}
 \caption{\small From (a) to (f), normal tumor image, and perturbed images with FGSM attack, with perturbation parameters: $\epsilon$ = 0.01, 0.03, 0.05,  0.10, respectively \peter{Also, it looks like this figure wasn't referenced in the text.} \quickthings{Added to the dicussion part b}}
 \label{fig:perturbedMRI}
\end{figure*}
\label{sec:results}

\label{sec:experiment}

\subsection{Datasets}
For our experiments, we employ non-independent and identically distributed (non-IID) samples across clients, which ensures that each client's dataset is distinct from the others. The datasets used in our study consist of a combination of multiple sources. This approach facilitates the assignment of diverse data to each client, thereby simulating real-world scenarios where data distribution is often non-uniform.

To allocate samples in non-IID manner, datasets were split into non-overlapping chunks, each chunk representing a tiny portion of the general distribution. Each client samples are taken only from specific chunks.

\textbf{Brain cancer classification:}  Dataset for detecting brain cancer is downloaded from Kaggle\cite{sartaj}. It contains brain MRI images categorized into four classes: three types of brain tumor,  and one healthy no Tumor class. 
For Meningioma and Glioma detection, 1437 and 1426 samples were used, respectively,  and they were split as train/test data. Transforms used are rotation, flipping, and normalizing.  Images were resized to 100$\times$100. 
\\
\textbf{Histopathologic cancer detection:} 
The samples are metastasic tissue images of lymph node cancer\cite{veeling2018rotation}. Each image has a resolution of 96 $\times$ 96 pixels, and the objective is to classify the tissue samples into cancerous and non-cancerous categories. In each positive sample, a metastatic region can be identified within a $32\times32$ pixel neighborhood, which is a square centered around the middle of the image, containing metastatic tissue.
A total of 2150 samples were selected and categorized into cancer and non-cancer classes. These samples were then randomly distributed among clients, with each client receiving a 62\% training split and a 38\% testing split.  We used horizontal flipping, and images were normalized.

\subsection{Network architectures}

\textbf{Deep learning models:}
The deep learning model used is a Convolutional Neural Network (CNN) with six layers of convolution stacked to 5 fully connected layers. The activation function used is ReLU and dropout parameter (0.25). Models are all trained and converged before implementing adversarial attacks.
For each dataset, a classifier is trained with Cross-Entropy loss and with an SGD optimizer.

\textbf{Federated setting:}
Three clients are defined for the federated setup. The data is assigned randomly, and the clients have non-IID data distribution.
The FedAVG method is used to aggregate the models. The aggregation is weighed based on the size of the training dataset. Each client is trained for 20 epochs at the communication round. The total federated rounds are 50.

\subsection{Attack setting}

The role of the adversarial client is alternated among the three clients, with each being assigned 100 test samples per use case for the attack. The averaged results of these attacks are evaluated using metrics like \textit{Clean Accuracy}, and \textit{Attack Success Rate}, which further leads to \textit{Average Attack Success Rate}.

We use the following metrics for evaluation:  \textit{Clean accuracy (ACC)} is defined as the performance of models on uncorrupted test images, \textit{Attack Success Rate (ASR)}; how much an adversary can change the predicted labels produced by each model. A formal definition of ASR is:
\begin{equation}
    \frac{1}{N}\sum\limits_{i=1}^{N} (Pre\text{-}attack\;label_{i}\neq Post\text{-}attack\;label_{i})
\end{equation}

Where N is the number of images in the attack set. We refer to the averaged ASR among all the clients as \textit{Average Attack Success Rate (AASR)}.



  Here, we expand the previous findings to the federated settings and discuss whether change in $\epsilon$ can lead to transferability. 
  
We perform three analyses to find important factors in attack success:
\begin{itemize}
    \item We investigate dependency of attacker on $\epsilon$. By visual inspection and ASR evaluation, and as a comparison to previously found optimal known values in centralized  setting.
    \item  We compare  efficiency of models, by discussing their attack preparation time and their ASR.
    \item We also see how attack step $\alpha$ can determine ASR.
\end{itemize}

\section{Results}






\subsection{Efficiency analysis}
Our results indicate that the computation has a linear relation with the attack power in PGD-like attacks, as shown in Table \ref{table_time}.  This is particularly important in considering the feasibility of attack methods in large-scale training of federated networks, and performing adversarial attacks on them efficiently\cite{chen2022configurable} \cite{chen2018bi}.
\begin{table}[h!]
\centering
\setlength{\tabcolsep}{6pt}
\renewcommand\arraystretch{1.22}
\caption{\small Comparison of iterative models and their computational efficiency, on performing computations on one batch of data. ACC shows average client performance for unperturbed test data. AASR is average ASR on all clients.}
\begin{tabular}{| *{5}{c|} }
\hline
Dataset  & Attack type & ACC & AASR & time (sec)
\\   \hline  
\multirow{2}{5em}{Meningioma}     &PGD-20&\multirow{2}{3em}{84.12\%}&27.52\% & 3.423 \\
&PGD-40&&32.99\% & 6.794  \\ \hline
\multirow{2}{4em}{Pathology}     &PGD-20&\multirow{2}{3em}{77.01\%}&60.98\% & 5.464\\
&PGD-40&&82.27\% & 10.843 \\ \hline
\multirow{2}{3em}{Glioma}     &PGD-20&\multirow{2}{3em}{61.84\%}&51.83\% &3.420  \\
&PGD-40&&63.77\% & 6.793  \\ \hline
\end{tabular}
\label{table_time} 
\end{table}

\subsection{Effect perturbation step}

Another critical parameter is the step of change in iterative methods ({$\alpha$}), which is shown to be highly deterministic in specific settings but is not yet investigated in medical image analysis domain. To evaluate how $\alpha$ can affect transferability, we examined the clients with $\epsilon =0.03$ in various steps. Generally, there is an optimal middle range the same as $\epsilon$. A step size of 0.007 led to the highest transferability in MRI images. Larger steps led to a sharp decrease in transferability. For pathology images, about 0.007 led to the best ASR on the adversary, and higher values decreased both ASR on adversary and transferability. However, smaller steps also had high ASR values. 

\begin{figure*}[h!]
    \centering
    \begin{subfigure}[b]{0.32\textwidth}
        \includegraphics[width=\textwidth]{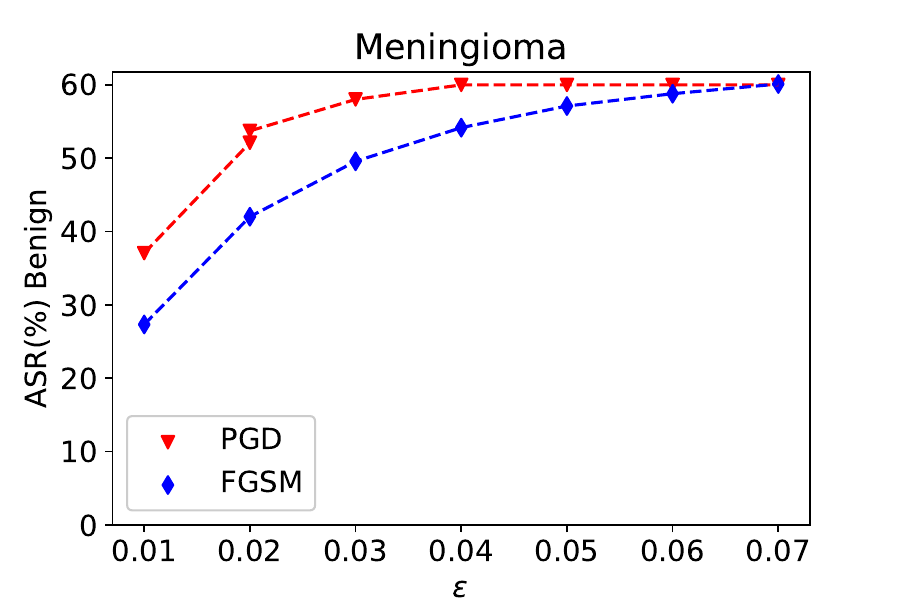}
        \caption{Meningioma benign }
        \label{fig:meningioma_asr}
    \end{subfigure}
    \begin{subfigure}[b]{0.32\textwidth}
        \includegraphics[width=\textwidth]{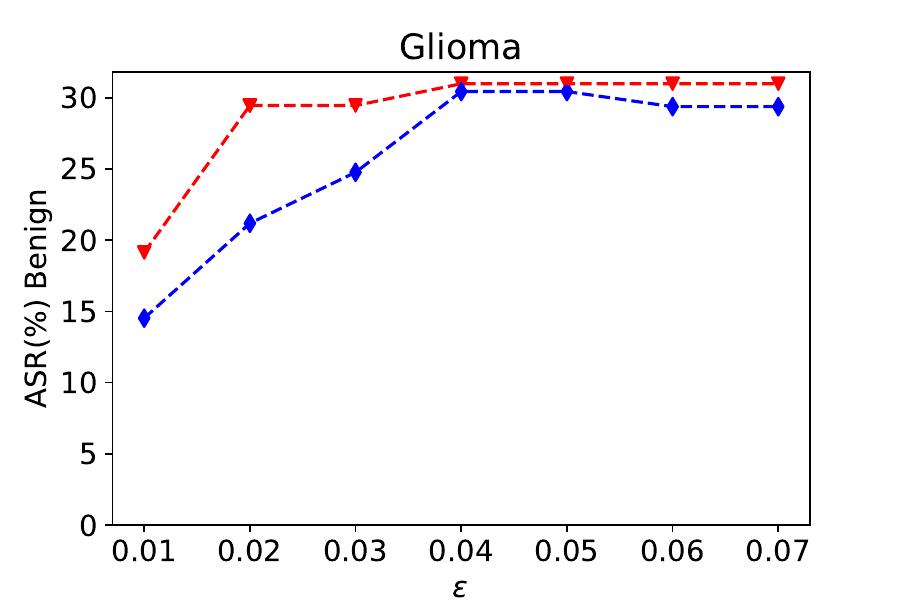}
        \caption{Glioma benign}
        \label{fig:glioma_asr}
    \end{subfigure}
    \begin{subfigure}[b]{0.32\textwidth}
        \includegraphics[width=\textwidth]{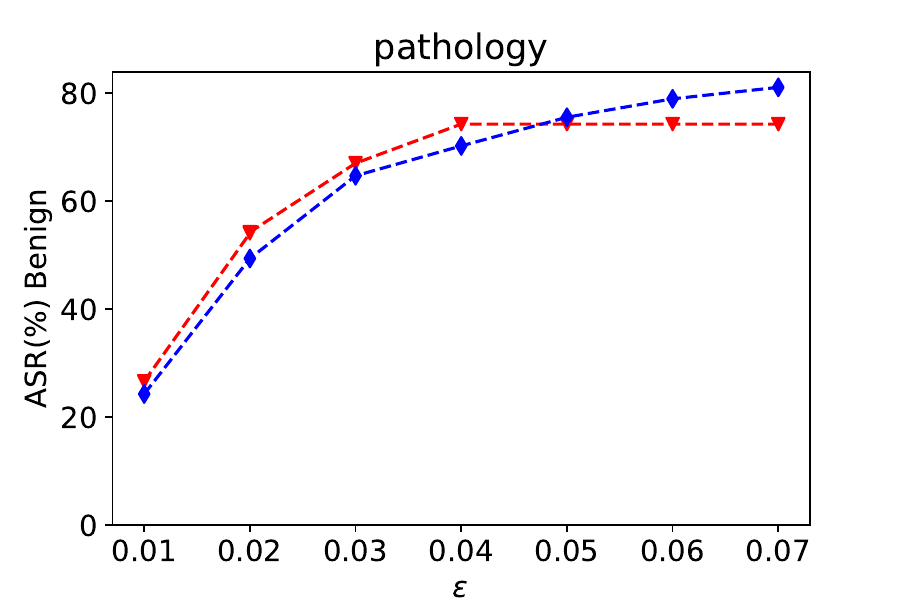}
        \caption{Pathology benign}
        \label{fig:pathology_asr}
    \end{subfigure}
    \begin{subfigure}[b]{0.32\textwidth}
        \includegraphics[width=\textwidth]{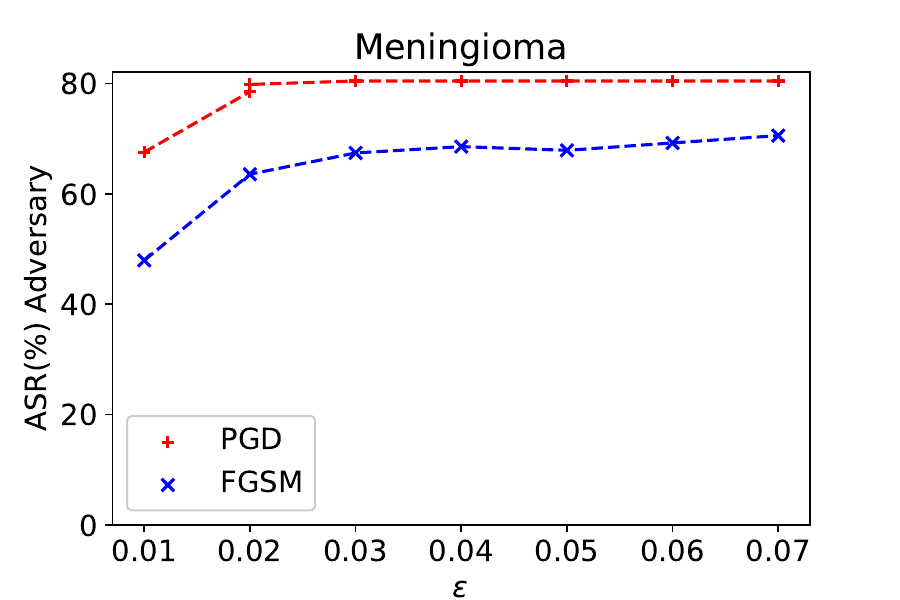}
        \caption{Meningioma adversary}
        \label{fig:meningioma_others}
    \end{subfigure}
    \begin{subfigure}[b]{0.32\textwidth}
        \includegraphics[width=\textwidth]{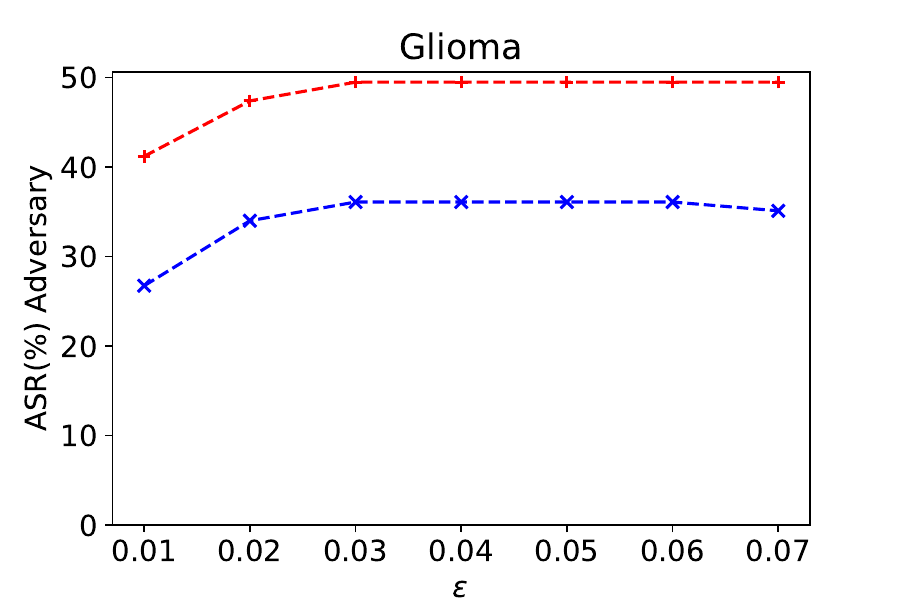}
        \caption{Glioma adversary}
        \label{fig:glioma_others}
    \end{subfigure}
    \begin{subfigure}[b]{0.32\textwidth}
        \includegraphics[width=\textwidth]{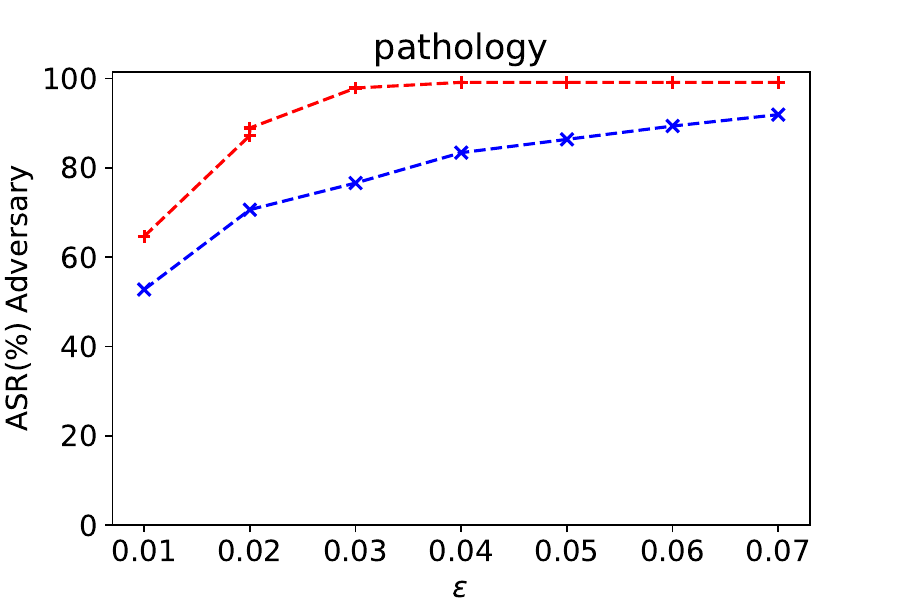}
        \caption{Pathology adversary}
        \label{fig:pathology_others}
    \end{subfigure}
    \caption{ASR vs. $\epsilon$ for different datasets. (a)-(c) show ASR for Meningioma, Glioma, and Pathology datasets, in setting where adversary attacks other clients, while (d)-(f) show the results for the adversarial client.}
    \label{fig:all_images}
\end{figure*}

\subsection{Effect perturbation degree}
Tuning $\epsilon$  might be tricky since performance and imperceptibility should be considered together.High $\epsilon$ values mean more distortion in the image. We performed an FGSM attack with varying $\epsilon$ values to evaluate the perturbation effect visually. Fig. \ref{fig:perturbedMRI} shows the results of perturbed images with different perturbation degrees and their comparison with the unperturbed image. Note that perturbation values below  $\epsilon=0.05$ cause limited distortion. Iterative models can reach high ASR very quickly with increasing $\epsilon$. Quantitative results of the effect of $\epsilon$ on attacker's success is shown in Figure \ref{fig:all_images}.

\begin{table*}[h!]
\centering
\setlength{\tabcolsep}{8pt}
\renewcommand\arraystretch{1.4}
\caption{Result of Average Error Transfer Rate (AETR) for FGSM and PGD methods. Each number is the average result for three experiments.}
\begin{tabular}{|c|c|c|}
\hline
Dataset & FGSM & PGD \\  
\hline
Meningioma & 84.80\% & 83.73\% \\
\hline
Glioma & 82.32\% & 74.89\% \\
\hline
Pathology & 90.67\% & 79.86\% \\
\hline
\end{tabular}
\label{table_AETR} 
\end{table*}


\section{Discussion}
\label{sec:discussion}
Our results confirm  prior research about vulnerability of FL networks
\cite{darzidehkalani2022federatedeco} and medical image analysis systems to adversarial attacks.


We also found that differential privacy might have partial effect on attacker's success. In fact, DP caused the lower ASR in benign than the adversary clients.
Although some research supports DP, our findings suggest that it is an option but not very reliable \cite{darzidehkalani2022federated2}.Additionally, it is important to consider the accuracy compromise that arises from the noise introduced by DP. This trade-off between privacy protection and model performance should be carefully evaluated, as excessive noise could negatively impact the model's overall accuracy and effectiveness in real-world applications.
\\In our experiments the attacker did not tamper with the training procedure. This is unlike backdoor or poisoning attacks , where the attacker performs malicious activities during training, and opens the way for detection methods.This section discusses our findings, how they could be important in the medical imaging domain, and what to consider when setting up federated medical image analysis infrastructure.

\subsection{Transferability}
Our experiments aimed to asses some factors in transferability. However, in addition to the investigated factors, we found three other parameters that might highly predict transferability.

\textbf{Benign/Adversary correlation:} Our results show that ASR on benign and adversary clients are highly correlated. Hence, for the adversary to ensure it can fool other clients in a black-box setting, it needs to obtain good results on its own data and model. So tuning the hyperparameters, data preprocessing can be all effective as long as they improve ASR on adversary.

\textbf{AETR/ASR difference:} As shown in Table \ref{table_AETR}, have higher AETR than ASR, which enables the adversary to subsample examples to achieve higher transferability. 

\textbf{Attack domain}: We observed that transferability is specific within imaging domains, which is consistent with prior findings. Attacks on Pathology images were consistently more successful than MRI images. 


\textbf{Step parameter:} Our findings show that common $\alpha$ values have a  negligible effect on the final results, aside from too small or too large values. The reason could be that $\alpha$ similar to  $\epsilon$  bounds the change in each iteration, but generally, its values are one order of magnitude less than $\epsilon$.

\subsection{Practical implications and suggestions}
Advanced fabrication algorithms and hard to detect methods are always intriguing for fraudsters. Existing Manipulations like altering the visual features of images, or using photo editing software,   can result in images of benign subjects classified as malignant,\peter{This should be discussed in the introduction. The motivation and goal for the attacker seems unclear.}\quickthings{I have moved some parts to the introduction and kepts some of it here. Hope it's okay now.}
\peter{This seems to be a goal also of backdoor attacks which I think are more concerning than pure model poisoning. It would be interesting to see if the federated setting somehow affects the trained global model when the adversarial noise it added at test time. Are there any differences to models that are trained in a centralized fashion?}
\quickthings{Interesting point of view. Since it is in the test time, it is fundamentally different than backdoor, poisonining , flipping etc. which all happen during the training.. The question of those attacks are " how can we change the model parameters that for a set of non manipulated test data, we have a high classification error" 
 The question here is " How can we change the input test data, that for a set of not manipulated models, we have a high classification error", so maybe a better experiment would be with a radnom noise?}but with visual distortion. In contrast, adversarial attacks are imperceptible, do not require manual intervention, and have high transferability. That gives a significant incentive for potential fraudsters to utilize adversarial attacks for high potential revenue. \\
Such vulnerabilities entail implications for clinical decision-makers, stakeholders, and insurance companies considering the deployment of federated or AI pipelines, which we discuss below:

\begin{enumerate}[(i)]
\item Manipulation and fraud, pervasive in healthcare and reinforced by financial and personal incentives, must be considered\cite{finlayson2019adversarial}.
\item A secure infrastructure during training does not guarantee safety during deployment; therefore, the distinction must be clear, and alternative methods may be required to integrate AI into medical image analysis pipelines.
\item Size and trustworthiness are crucial parameters in collaborative/federated networks. Smaller networks with trusted parties reduce the likelihood of adversaries.
\item Employing diverse data and not enforcing uniform data pre-processing pipelines impedes adversaries. However, disparate data may compromise performance or convergence of federated networks , necessitating healthcare managers to weigh security against performance.
\end{enumerate}

Developers of medical image analysis systems should consider the following:

\begin{enumerate}[(i)]
\item Providing end-users and clinicians with supplementary information, such as explainable system reports, to evaluate prediction legitimacy.
\item Employing suggestions fromto test adversarial robustness in centralized settings is inadvisable in federated settings. Developers must consider scenarios where adversaries trace the global model and employ unexpected devices.
\item Attack settings and parameters like $\epsilon$ affect transferability. Adversaries may enhance attacks by selecting suitable settings. Developers should utilize prior research findings to determine optimal parameters for their domain or conduct brute-force searches to identify vulnerability upper bounds .
\item Although no universal defense method exists, limited case-specific defense models have proven effective on select datasets , warranting exploration of potential defense methods for specific use cases.
\end{enumerate}

\peter{Adding some discussion on how the attack could be detected would be good.}\quickthings{Added!}
Future lines of research could be to improve existing defense methods , towards a universal defense algorithm. Although there are some promising results, Some of them are only usable under specific settings,like they work for manipulations that cause a pattern that is previously known by the defender. Some of the algorithms require too much computational power, as they need to assume multiple scenarios and retrain the model on all of the corrupted samples generated from all those scenarios. Some detection models lead to drastic model parameter change. Also, some recent studies have shown that the existing defense methods are ineffective if the adversary is aware of them . 

Another line of research could be how to utilize distributed nature of federated networks to protect all participating clients, or if collaboration can be used to enhance the current defense methods.\\


\section{Conclusion}
\label{sec:conclusion}

This paper investigated adversarial attacks on federated medical image analysis systems and discussed the crucial parameters on their transferability.
Our results highlight the potential misuse of adversarial attacks and emphasizes the need for defense mechanisms. Clinical decision-makers, stakeholders, and insurance companies are urged to be aware of these risks when deploying AI or federated pipelines. The research indicates that security during training does not ensure safety during deployment, necessitating alternative strategies for AI integration in medical image analysis pipelines. Developers are encouraged to consider various adversary scenarios and continue exploring potential defense methods. The paper also suggests future research towards a universal defense algorithm. In conclusion, this research offers insights into understanding adversarial attacks in federated medical image analysis systems and aims to equip healthcare institutions with the knowledge to mitigate these threats.

\ifCLASSOPTIONcaptionsoff
  \newpage
\fi





\bibliographystyle{IEEEtran}
\bibliography{IEEEabrv}

\vfill

\end{document}